\documentclass[prb,twocolumn,showpacs,amsmath,amssymb,preprintnumbers]{revtex4}
\usepackage{dcolumn}
\usepackage{bm}
\usepackage{graphicx}
\usepackage{color}

\begin{document}

\title{Insight into the magnetic behavior of Sr$_2$IrO$_4$:A spontaneous magnetization study}

\author{Imtiaz Noor Bhatti}\email{inbhatti07@gmail.com}\affiliation{School of Physical Sciences, Jawaharlal Nehru University, New Delhi - 110067, India.}
\author{A. K. Pramanik}\affiliation{School of Physical Sciences, Jawaharlal Nehru University, New Delhi - 110067, India.}

\begin{abstract}
Sr$_2$IrO$_4$ is a weak ferromagnet where the spin arrangement is canted anti-ferromagnetic (AF). Moreover, the spin-structure coupling plays and important role in magnetic behavior of Sr$_2$IrO$_4$. In this concern the magnetization under zero applied field i.e. spontaneous magnetization would be interesting to study and would give insight into the novel magnetic behavior of Sr$_2$IrO$_4$. Sophisticated techniques like neutron diffraction, $\mu$ \textit{SR} etc has been used to understand the magnetic behavior of Sr$_2$IrO$_4$ under zero applied field. To understand the magnetic behavior we have performed detail field and temperature dependent magnetization study, the measured field and temperature dependent magnetic data is analyzed rigorously. We have attempted the understand the temperature dependance of spontaneous magnetization, remanent magnetization and coercive force. We observe that the spontaneous magnetization extracted from Arrott plot shows that the Sr$_2$IrO$_4$ is not an ideal ferromagnet. The temperature dependent coercive field is found to follows Guant's model of strong domain wall pinning.  Our investigation explicit the temperature dependence of various magnetic properties shows the magnetic transitions from paramagnetic to ferromagnetic phase with $T_c$ around 225 K and a low temperature evolution of magnetic magnetic moment around $T_M$ $\sim$90 K.
\end{abstract}

\maketitle 

\section{Introduction}
The magnetic ground state of layered Sr$_2$IrO$_4$ is canted anti-ferromagnetic with phase transition $\sim$ 225 K driven by spin orbital interaction and lattice distortion.\cite{crawford, imtiaz1, ye, moon, gao, yan} In this case the spins are not exactly anti-parallel but tilted a an angle to its relative spins called spin canting behavior. The canted moment is resulted from antisymmetric Dzyloshinskii-Moriya ($DM$) interaction induced through the interaction between large SOC and lattice distortion. The large canted moment imply strong anisotropy in this material.\cite{crawford} Moreover, x-ray resonant magnetic scattering study on  Sr$_2$IrO$_4$ establish the  basal-plane antiferromagnetic structure with spin canting by $\sim$11$^o$.\cite{kim} Sr$_2$IrO$_4$ is first member of Ruddelsden-Popper series  Sr$_{n+1}$Ir$_n$O$_{3n-1}$ with n=1 is  a single layered perovskite structure with reduced tetragonal symmetry due to IrO$_6$ octahedral rotation about \textit{c}-axis $\sim$ 11$^o$. The IrO$_6$ octahedral distortion  give special feature to structure organization of Sr$_2$IrO$_4$. Moreover the structural parameters evolve with temperature and are linked with magnetic transitions.\cite{imtiaz1} Since the spin are sitting in the center of IrO$_6$ octahedral thus staggered rotation of spins along with octahedron are expected.  It is this spin canted nature of magnetic ordering which gives spontaneous magnetization and hence weak ferromagnetic character to Sr$_2$IrO$_4$. Thus it would be quite interesting to investigate the spontaneous moment.Indeed it is this deviation which is the source of spontaneous magnetization conformed by advanced techniques like neutron diffraction and $\mu$ \textit{SR}.\cite{ye}

	This material isoelectronic and isostructure to La$_2$CuO$_4$ thus received attention in recent time due to predicted superconductivity and exotic properties.\cite{moon, gao, yan, tarascon,bozin} The investigation for magnetism in Sr$_2$IrO$_4$ have been reported, both theoretical and experimental attempts are made to understand the two-dimensional  system. Recently it has been shown that introduction of Hund's coupling correction gives anisotropic terms in Hamilttonian. The J$_{eff}$ moment is suppose to be strongly coupled to background lattice, a stagger rotation of spins with IrO$_6$ octahedral is predicted.\cite{jackeli} Further,  investigation using fully relativistic constrained non-collinear density functional theory and clarify that the canted nature of magnetic ordering in Sr$_2$IrO$_4$ is resulted from structural distortion along with interplay of  isotropic exchange and antisymmetric $DM$ interaction.\cite{liu} Recently, longitudinal and torque magnetometry is employed to study magnetic anisotropy in Sr$_2$IrO$_4$. Basel-plane magnetocrystlline anisotropy is accounted for deviation of magnetization direction from appled field.\cite{fruchter} Anisotropic exchange interaction has been discussed in detail. Experimentally, strong magnetic anisotropy between in plane and out of plane direction is observed, which persist upon chemical doping.\cite{fujiyama, ge} Signature of large magnetic anisotropy between the directions i.e. \textit{ab} plan and \textit{c} axis is observed in magnetization study.\cite{cao} Recently it is found that the Tb doping strongly affect the magnetic anisotropy which is expected from strong 5d-4f interaction.\cite{wang} In our earlier work we found that there is coupled evolution of structure and magnetism in Sr$_2$IrO$_4$.\cite{imtiaz} Recently, constrained density functional theory investigation have revailed anisotropic magnetic couplings and structure-driven canted to collinear transitions in Sr$_2$IrO$_4$.\cite{sharma}
Further, investigation using resonant inelastic X-ray scattering (RIXS), X-ray magnetic critical scattering also establishes the easy-plane anisotropy present in Sr$_2$IrO$_4$. \cite{vale, bahr} The pseudodipolar interactions resulted from the intrinsic spin-orbit coupling and spin structural coupling is believed to give rise to this anisotropy.\cite{liu, porras} Sr$_2$IrO$_4$ is an canted antiferromagnetic material which give rise to a weak ferromagnetic component in this material with T$_c$ $\sim$226 K. Domain structural and domain wall is further a interesting aspect to study in this compound. Some efforts have been made to understand the domail wall structure in such compounds with weak ferromagnetic nature.\cite{li, hir} In this paper we have attampted to shade some light in this direction in weak ferromagnetic Sr$_2$IrO$_4$.
	
	\begin{figure}
\centering
	\includegraphics[width=8cm]{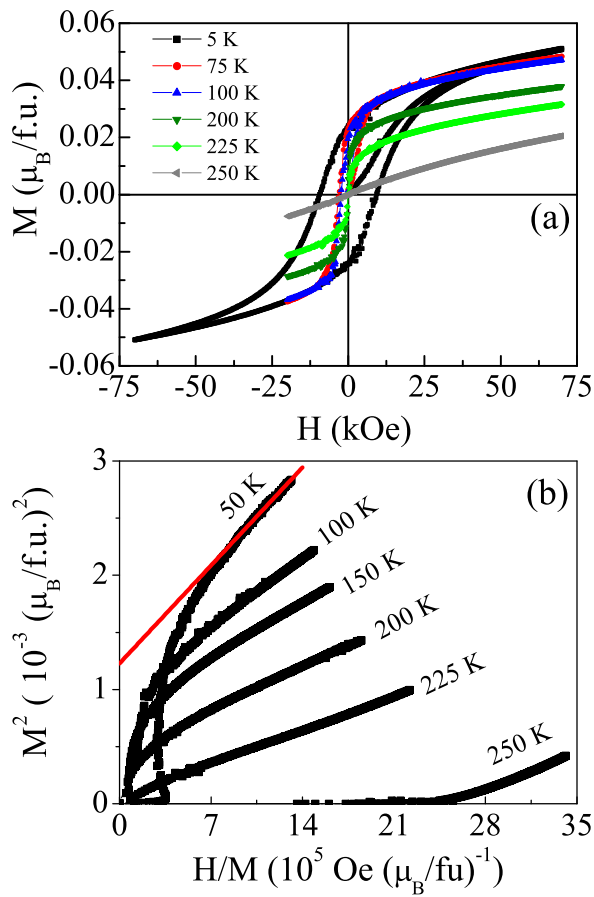}
\caption{(color online) (a) Four representative $M(H)$ curves collected at different temperature for Sr$_2$IrO$_4$. (b) Arrot plot $M^2$ vs $H/M$ plotted using virgin curves of field dependant magnetization $M(H)$ at different temperature.}
\label{fig:Fig1}
\end{figure}
	
	In this study we present temperature dependent magnetic properties of Sr$_2$IrO$_4$ in detail. Evolution of spontaneous magnetization with temperature is studied, which shows that the magnetic behavior is not ideal ferromagnetic but at low temperature favor anti-ferromagnetic ordering.  Temperature dependence of coercive force and remanence is studied and shows strong domain wall pinning (SDWP) scenario for Sr$_2$IrO$_4$.

\section{Experimental Detail}
Single phase polycrystalline sample was prepared by standard solid state reaction method, the detail procedure of sample preparation and phase purity by X-ray diffraction is reported elsewhere.\cite{imtiaz, imtiaz1, imtiaz2} Field dependent magnetization $M(H)$ has been measured at different temperature from 5 K to 300 K at temperature interval of $\Delta$$T$ = 25 K. Special care is taken in data measurement, before every $M(H)$ scan sample is warmed above 250 K to its paramagnetic phase and field ($H$) set to zero in oscillate mode to nullify the influence of high field on spin orientation. The magnetization data is recorded using physical property measurement system (PPMS) by Quantum Design.

\section{Result and Discussion}
The ferromagnetic component in Sr$_2$IrO$_4$ is a result of canted nature of spin lattice. Neutron diffraction studies show that the canted type anti-ferromagnetic arrangement of spins, is resulted from lattice distortion along with strong spin orbital coupling.\cite{ye} The spin and IrO$_6$ octahedral distortion are coupled and straggly rotate along \textit{c}-axis.\cite{jackeli} Indeed, we have shown that the octahedral distortion and spin canting  evolve with temperature which corroborates with the evolution of magnetic moment with temperature.\cite{imtiaz} Similar, results have also been reported in temperature dependent Raman study in Sr$_2$IrO$_4$.  It is this canting nature of spin which gives spontaneous moment and ferromagnetic character in Sr$_2$IrO$_4$.\cite{crawford} Thus, investigation of spontaneous magnetization can be interesting and would help to further understand the magnetic behavior in Sr$_2$IrO$_4$.

In order to investigate the magnetic behavior of Sr$_2$IrO$_4$ we have performed detail magnetization measurement. Fig. 1(a) shows representative isothermal magnetization $M(H)$ recorded at some selective temperatures. For detail study, we have measured the $M(H)$ data in the range of 5 K to 300 K with temperature interval of $\Delta$$T$ = 25 K. The $M(H)$ plots in Fig 1(a) clearly show hysteresis which conforms the existence of  ferromagnetic component in Sr$_2$IrO$_4$. Ferromagnetic materials exhibits a net magnetic moment in the absence of an external magnetic field called spontaneous magnetization ($M_S$). Arrott plot ($M^2$ vs $H/M$) is an effective tool to extract the spontaneous magnetization, where plots at the high field are extrapolated and there intercept at $M^2$ axis will give $M_S$. Using the $M(H)$ data from Fig. 1(a) for Sr$_2$IrO$_4$, we have plotted Arrott plot as shown in Fig. 1(b). These, Arrott plots were used to extract the $M_S$ shown in Fig. 2(b). 

\begin{figure}[th]
\centering
		\includegraphics[width=8cm]{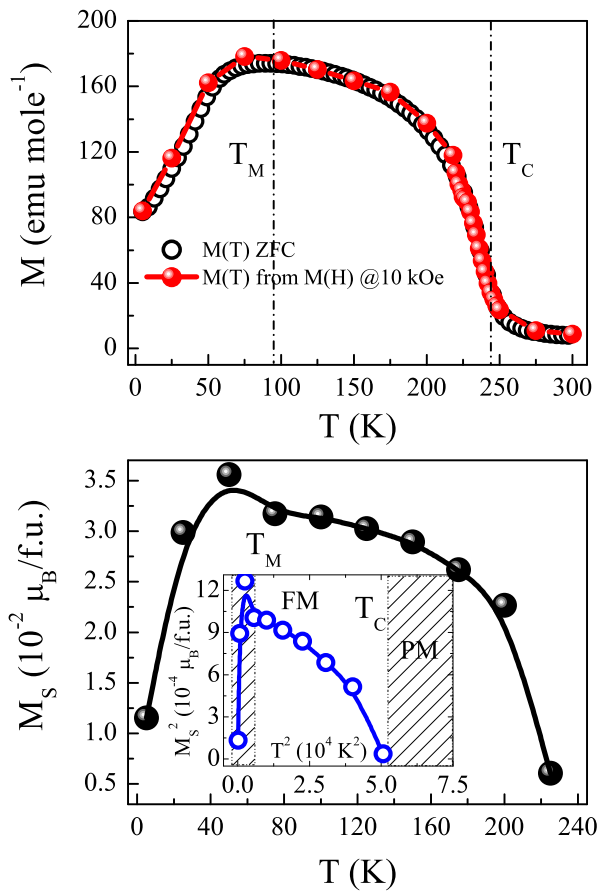}
	\caption{(color online) (a) Comparative plot of temperature dependent magnetization $M(T)$ measured under 10 kOe of external applied DC magnetic field on cooling in zero field and $M(T)$ extracted from $M(H)$ virgin curves at same field  for Sr$_2$IrO$_4$. (b) Temperature dependance of spontaneous magnetization $M_S$ extracted from the intercept on y-axis of high field fitting of Arrott plots for Sr$_2$IrO$_4$ in Fig. 1(b). Inset shows $M{_S}{^2}$ vs $T^2$ plot.}
	\label{fig:Fig2}
\end{figure}

\begin{figure}[th]
\centering
	\includegraphics[width=8cm]{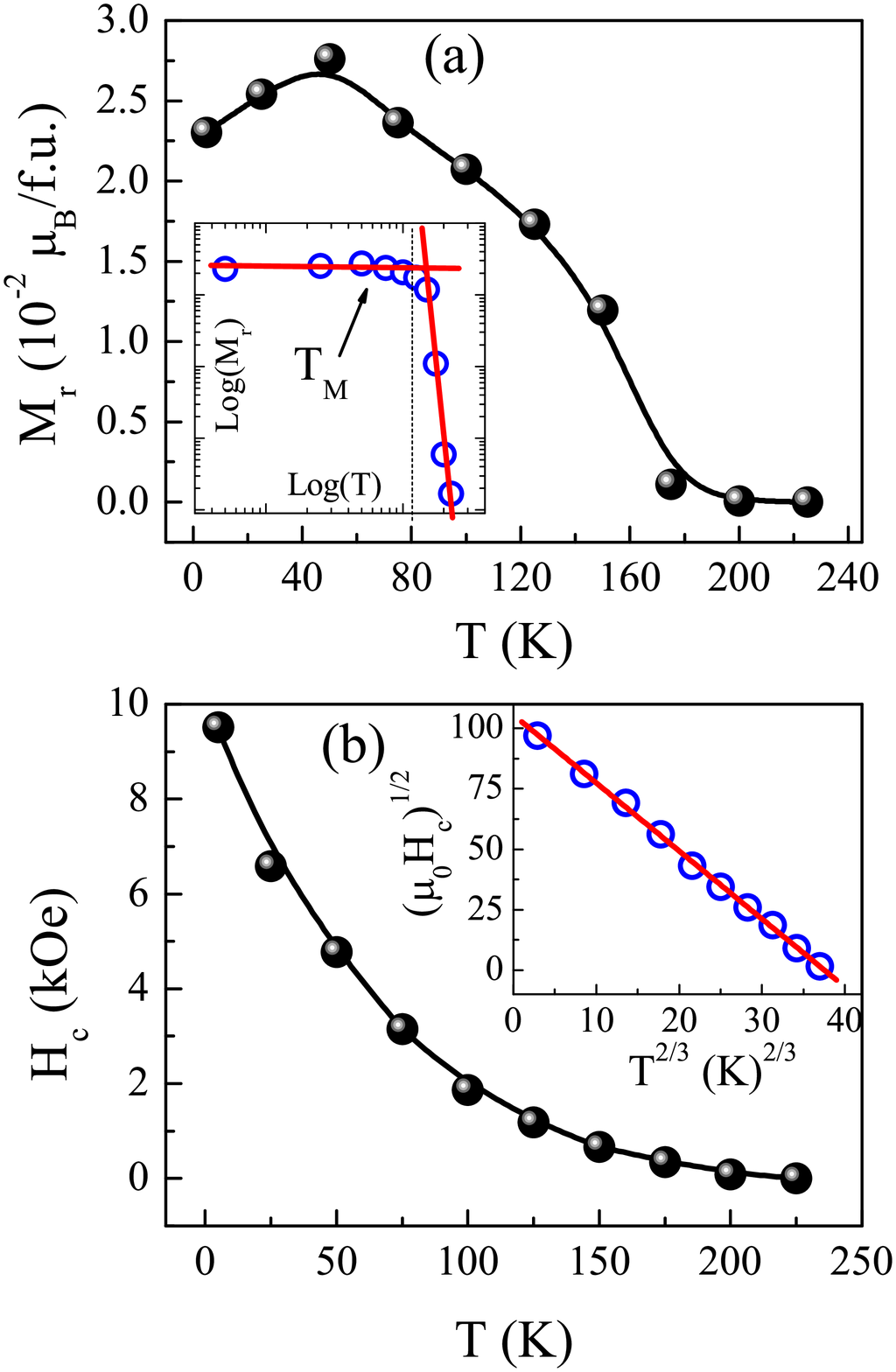}
	\caption{(color online) (a) Temperature dependence of remanent magnetization ($M_r$($H$ = 0) vs $T$) of Sr$_2$IrO$_4$. Inset log-log plot of $M_r$($H$ = 0) vs $T$ shows two regions above and below $T_M$ $\sim$ 90 K. (b) Temperature dependent  coercive field of Sr$_2$IrO$_4$.Inset shows $H{_c}^{1/2}$ vs $T^{2/3}$ showing the linear behavior expected in the SDWP model, solid lines is fit due to equation 1.}
	\label{fig:Fig3}
\end{figure}

Fig. 2(a) (open black circles) shows the temperature dependent magnetization $M(T)$ measured in zero field cooled protocol ($ZFC$) at applied field of 10 kOe.We have further determined the $M(T)$ data by extracted from $M(H)$ data at 10 kOe Field shown as blue bold circles in Fig. 2(a). The moment at 10 kOe extracted from $M(H)$ data is exactly follow the same trend as $M(T)$ does. It is clear from Fig. 2(a) that Sr$_2$IrO$_4$ shows evolution of magnetization with temperature a paramagnetic ($PM$) to ferromagnetic ($FM$) phase transition around $T_c$ = 225 K and a low temperature downfall in magnetic moment marked my $T_M$ around 90 K after which moment start decreasing dotted lines are guide to eyes. The downfall in magnetic moment at low temperature is believed to be resulted from reduced canting of spins and favor anti-ferromagnetic ordering. We have extracted the spontaneous moment $M_S$ at selective temperatures ($T$) from intercept on positive $M^2$ axis  by extrapolating the high field region of Arrott plot ($M^2$ vs $H/M$) shown in Fig. 1(a). Temperature dependence of $M_S$ shown in Fig. 2(b) reveals that the ferromagnetic component in Sr$_2$IrO$_4$ changes with temperature. Since the magnetic ground state of Sr$_2$IrO$_4$ is canted antiferromagnetic resulted from $DM$ interaction produced by spin orbital coupling and structural distortion. Inset Fig. 2(b) shows quadratic spontaneous magnetization ($M{_S}{^2}(T)$) plotted against quadratic temperature ($T^2$). The  $M{_S}{^2}(T)$ increase with decreasing temperature but not much deviated from linearity which is understood because recently it is shown that Sr$_2$Ir$O_4$ can be a case with  mixture of both localized and itinerant mechanism.\cite{imtiaz1} Moreover, the $M_S$ shows a strong deviation around $T_M$ $\sim$ 90 K which is understood since degree of spin canting is reduced and AFM ordering favor at low temperature.
	Fig. 3(a) shows temperature dependence of remanent magnetization ($M_r$), with decreasing temperature $M_r$ increases and attain a maximum value at temperature $\sim$ 50 K and start decreasing with further lowering temperature. However, inset Fig. 3(a) shows log-log plot of remanent magnetization vs temperature which clearly depicts two regimes one below and above $T_M$. This is already reported that the low temperature  evolution of magnetic phase at $T_M$ is coupled with structure modification where $<$Ir-O-Ir$>$ bond angel tends to straighten and octahedral rotation $\theta_{Oct}$ reduces. This favors antiferromagnetic ordering and hence reduces the net magnetic moment.\cite{imtiaz}
	The magnetic coercivity  is the ability of magnetic material to withstand external magnetic field without being demagnetized. The $M(H)$ curves for Sr$_2$IrO$_4$  (Fig. 1(a)) are used to extract the coercive field and the values are plotted against temperature in Fig. 3(b). We have shown temperature dependent coercive field inset Fig. 3(b), the coercivity increase with decreasing temperature. The temperature dependence of coercive field obeys the strong pinning of ferromagnetic domains wall obeys the model proposed by Guant.\cite{guant}
\begin{eqnarray}
\mu_0H{_c}^{1/2} = 1 - \left[\frac{75kT}{4\textit{bf}}\right]^{2/3}
\end{eqnarray}
where H$_c$ is coercive field, \textit{f} is the magnetic force required to depin a domain wall, 4\textit{b} is domain wall width and k$_B$ is Boltzmann constant. from the fitting perameter we obtain the value of 4$bf$ = 1.589 $\times$ 10$^{-15}$ erg. The variation of coercive field is described by SDWP model throughout the temperature range.

\section{Conclusion}
In conclusion, ground state in Sr$_2$IrO$_4$ is canted anti-ferromagnetic where spin are aligned to some degree instead of being exactly anti-parallel which gives ferromagnetic component found to varies with temperature. We found that the spontaneous magnetization evolve with temperature and shows sudden decrease below 100 K which is evident that at low temperature the degree of spin canting with temperature decreases and more favor to anti-parallel spin arrangement. The remanent magnetization increase with temperature upto 100 K and decrease afterward which is understood due to reduced spin canting and favor AFM ordering. Further, the temperature dependent coercivity can be best fitted to Guant model which predict strong pinning due to magnetic inhomogeneities.

\section{Acknowledgement}
We acknowledge UGC-DAE CSR, Indore for magnetization data.

\end{document}